\documentclass[aps.pra,twocolumn,showpacs,preprintnumbers,amsmath,amssymb]{revtex4}
\usepackage{mathrsfs}
\usepackage{graphicx}
\usepackage{amsmath}
\usepackage{ulem}
\usepackage{color}
\newcommand{\vect}[1]{\vec{#1}}

\begin{document}

\title{Dynamical Lamb effect versus dissipation in superconducting quantum circuits}
\author{A. A. Zhukov$^{1,3}$, D. S. Shapiro$^{1,2}$, W. V. Pogosov$^{1,4}$, Yu. E. Lozovik$^{1,5,6}$}

\affiliation{$^1$N. L. Dukhov All-Russia Research Institute of Automatics, 127055 Moscow, Russia}
\affiliation{$^2$V. A. Kotel'nikov Institute of Radio Engineering and Electronics, Russian Academy of Sciences, 125009 Moscow, Russia}
\affiliation{$^3$National Research Nuclear University (MEPhI), 115409 Moscow, Russia}
\affiliation{$^4$Institute for Theoretical and Applied Electrodynamics, Russian Academy of
Sciences, 125412 Moscow, Russia}
\affiliation{$^5$Institute of Spectroscopy, Russian Academy of Sciences, 142190 Moscow region,
Troitsk, Russia}
\affiliation{$^6$Moscow Institute of Electronics and Mathematics, National Research University Higher School of Economics, 101000 Moscow, Russia}

\begin{abstract}
Superconducting circuits provide a new platform to study nonstationary cavity QED phenomena.
An example of such a phenomenon is a dynamical Lamb effect which is a parametric excitation of an atom due to the nonadiabatic
modulation of its Lamb shift. This effect has been initially introduced for a natural atom in a varying cavity, while we suggested its realization in a superconducting qubit-cavity system with dynamically tunable coupling. In the present paper, we study the interplay between the dynamical Lamb effect and the energy dissipation, which is unavoidable in realistic systems. We find that despite of naive expectations this interplay can lead to unexpected dynamical regimes. One of the most striking results is that photon generation from vacuum can be strongly enhanced due to the qubit relaxation, which opens a new channel for such a process. We also show that dissipation in the cavity can increase the qubit excited state population. Our results can be used for the experimental observation and investigation of the dynamical Lamb effect and accompanying quantum effects.
\end{abstract}

\author{}
\maketitle
\date{\today }

\section{Introduction}

Superconducting circuits can be exploited for experimental
investigation of cavity quantum electrodynamical (QED) effects
\cite{Nation}. This possibility is due to the recent progress in
fabrication methods and quantum fields control which allows one to
use superconducting systems in quantum information and computation
\cite{Devoret, Astafiev, Oelsner, Martinis, IBM, Ustinov}. The
transfer of information can be efficiently implemented provided
dissipation effects and external noise are ruled out, while this
problem is known to be quite difficult to solve \cite{Makhlin}.
Obviously, similar requirements have to be fulfilled for QED effects
to be observed in experiments.

Several years ago a first observation of one of the most intriguing
nonstationary QED phenomena known as the dynamical Casimir effect
was reported \cite{DCE1, DCE2}. It is quite remarkable that it was
observed for the first time in superconducting systems although it
has been predicted for systems seemingly very distinct from such
circuits \cite{Moore}. Being nonstationary, the dynamical Casimir
effect differs from static Casimir effect. Static Casimir effect is
manifested as an attraction of two static mirrors due to the
zero-point fluctuations of photon field confined between them. These
vacuum fluctuations contribute also to another well known static QED
effect -- Lamb shift of atom spectrum. Such a shift exists not only
for natural atoms, but also for artificial superconducting "atoms"
(qubits) coupled to resonators \cite{Lamb}. Moreover, in contrast to
natural atoms, the effect can be significantly enhanced, since a
regime of strong qubit-cavity coupling is achievable in such
systems.

Dynamical Casimir effect was initially predicted to occur provided mirrors are moving with respect to each other. This motion leads to a modulation of allowed photon wave vectors, as dictated by quantization conditions. As a result, real photons are generated from a vacuum between the mirrors thus parametrically amplifying vacuum fluctuations. In order to produce a feasible photon emission rate, one has to move mirrors with velocities approaching a speed of light. For massive mirrors, this requirement is challenging for nowadays experimental facilities. That is why various indirect schemes have been suggested \cite{Yablonovitch,Lozovik-plasma,Dodonov,Braggio}, among which we mention modulation of electromagnetic properties of the cavity walls, the usage of acoustic waves and nanomechanical resonators. One of such proposals was to modulate an inductance of the superconducting quantum interference device (SQUID) connected to the coplanar waveguide \cite{Segev}. Such a modulation can be treated as a change of the electrical length of the waveguide which is accompanied by a desired variation of boundary conditions. Since no motion of massive objects is involved, a very fast modulation rate can be achieved.

We wish to stress that the dynamical Casimir effect is just one of
the examples of a large class of nonstationary QED effects in which
vacuum amplification is expected to play a major role \cite{Nation}.
The most known effects of this kind are the Unruh effect
\cite{Unruh} and the Hawking radiation \cite{Hawking}. None of these
phenomena has been observed so far except of the dynamical Casimir
effect.

A very rich behavior is also demonstrated by nonstationary cavities containing a single atom or an ensemble of such atoms thanks to the matter-light coupling, see, e.g., Refs. \cite{Hein,Pokrov,Lozlett,Lozovik1,review,VDodonov,ADodonov,ADodonov1}. The simplest possible system of such kind is a single-mode cavity with time dependent frequency, which contains a two-level system. A cavity with nonadiabatically modulated frequency is similar to a parametrically driven harmonic oscillator. Such a modulation leads to the generation of Casimir photons, which can naturally be absorbed by the atom resulting in its excitation \cite{Lozlett}. A precise analysis \cite{Lozovik1}, however, shows that there is another channel of atom excitation which is due to the nonadiabatic modulation of its Lamb shift. This effect is related to the static Lamb shift in a similar way, as the dynamical Casimir effect is related to the static Casimir effect. For this reason, it was suggested in Ref. \cite{Lozovik1} to term this phenomenon as the "dynamical Lamb effect".

Unfortunately, experimentally it seems to be quite difficult to isolate the channel of atom excitation due to the absorption of Casimir photons from the mechanism due to the dynamical Lamb effect, since these two excitation channels always appear together in experiments with nonstationary cavities, in which their frequencies experience external variations.

Fortunately, instead of real atoms it is possible to use artificial "atoms" made of superconducting circuits with Josephson junctions. This opportunity is very attractive because of the high flexibility of such circuits. Two-level superconducting artificial "atoms" are used nowadays as qubits for purposes of quantum computation. Being macroscopic in their sizes, they can demonstrate quantum behavior on rather long time scales, approaching hundreds of microseconds for state-of-art devices. Quality factors of available microwave resonators are of the order of $10^6$, so that a coupled qubit-cavity system can behave quantum mechanically during time intervals needed to perform hundreds or thousands of quantum gates \cite{Martinis}.

Moreover, it is known that coupled systems of superconducting resonators and qubits can be fabricated as {\it dynamically} tunable {\it in-situ } during experiments. It was demonstrated that not only the resonator frequency and the qubit excitation energy can be modulated but one can also change a vacuum Rabi frequency determined by the strength of a qubit-resonator coupling. Such a modulation can be achieved using either flux qubits with an additional SQIUD or two strongly coupled charge qubits (transmons) out of which a single effective two-level system can be created, see, e.g., Refs. \cite{tunable1,tunable2,tunable3}. Thus it is possible not only to change several parameters simultaneously by perturbing the whole system, but also to modulate a particular {\it single} parameter.

This remarkable opportunity opens a possibility for the full isolation of the mechanism of qubit excitation due to the dynamical Lamb effect from the channel of its excitation due to the absorption of Casimir photons, as suggested recently in Ref. \cite{paper1}. Indeed, if one modulates only the qubit-resonator coupling and does not change a resonator frequency, no Casimir photons appear. Nevertheless, a qubit can be parametrically excited since it somehow "feels" a nonadiabatic change of its Lamb shift. In order to enhance the effect and to increase qubit excited state population, it was suggested in \cite{paper1} to modulate the resonator-qubit coupling periodically with twice the resonator frequency, while qubit and resonator are in a full resonance. However, in \cite{paper1} a dissipation in the qubit-resonator coupled system was completely ignored.

The major aim of the present paper is to treat an interplay between the dynamical Lamb effect and the dissipation within the realization proposed in Ref. \cite{paper1}.
The naive expectation is that dissipation must always suppress this purely quantum effect,
as well as the process of photon generation from vacuum. In particular, relaxation in qubit is
opposite to the qubit excitation process, induced by the dynamical Lamb effect, since it leads
to a qubit de-excitation. In reality, we find that the effect of dissipation is far more
complex and it results in several highly unexpected dynamical regimes including enhanced generation of photons from vacuum. Another regime resembles a parametric down conversion since it results in a generation of photon pairs with the frequencies lower than the pump frequency.

This paper is organized as follows. In Section II, we describe the system under consideration and
outline our theoretical model. In Section III, we present a simple toy model which allows us to understand some important features of the dynamical behavior of our system without performing
numerical simulations and under the assumption that the decay rate in the cavity can be neglected. The results of such simulations under the same assumption are then presented in Section IV. Section V deals with the analysis of the effect of cavity relaxation. In Section VI we apply an alternative method to solve a problem applicable for the stationary limit after the stabilization in order to crosscheck our main results. We conclude in Section VII.

\section{Hamiltonian and basic equations}

We are focused on superconducting circuits which consist of flux or charge qubit (transmon). This two-level system  is coupled dynamically to  high quality coplanar waveguide, playing a role of single mode cavity in optical systems. Qubit and waveguide  are spatially separated on a chip, indeed, coupling between them can be organized by auxiliary SQUID or by means of other methods \cite{tunable1,tunable3}. Current superconducting technologies allow to realize architectures where qubit-cavity coupling can be switched on and off or tuned  at GHz frequencies, while the amplitude  can be varied at values up to $100$ MHz. As can be expected, near-future technologies will be able to impose even more adjustable modulations.

We describe the photon mode and qubit (at frequencies $\omega$ and  $\epsilon$ respectively which are of the order of several GHz)  by means of the Rabi
model \cite{Rabi,JCmodel}, known from quantum optics, taking into account dynamically tunable coupling energy $g(t)$. Total Hamiltonian of this system reads
\begin{equation}
H(t)=\omega a^{\dagger }a + \frac{1}{2} \epsilon (1+\sigma_{3})+V(t),
\label{Hamiltonian}
\end{equation}
where $a^{\dagger }$ and $a$ are secondary quantized photon creation and annihilation operators and
$\sigma_{3}=2\sigma_{+}\sigma_{-}-1$, $\sigma_{+}$, $\sigma_{-}$ are Pauli operators related to qubit degrees of freedom. The non-stationary operator  $V(t)$ describes dynamical qubit-cavity coupling
\begin{equation}
V(t)=g(t)(a+a^{\dagger })(\sigma_{-}+\sigma_{+}),
\label{FullV}
\end{equation}
where $(a+a^{\dagger })$ and $(\sigma_{-}+\sigma_{+})$ are related to electric field and dipole
moment operators, respectively.

This qubit-cavity interaction operator can be divided into two parts
\begin{equation}
V(t)=V_1(t)+V_2(t),
\label{split}
\end{equation}
where
\begin{equation}
V_1(t)= g(t)(a \sigma_{+} + a^{\dagger } \sigma_{-})
\label{V1}
\end{equation}
is responsible for the well known rotating wave approximation (RWA), provided $V_2$ is dropped, while $V_2$ is given by
\begin{equation}
V_2(t)=g(t)(a^{\dagger } \sigma_{+} + a \sigma_{-}).
\label{V2}
\end{equation}
This counter-rotating wave term (CRT) is usually neglected.

The term $V_1$ in the Hamiltonian conserves total number of excitations in the system and in stationary case $V_1(t)=\rm{const}$ it provides exactly solvable Jaynes-Cummings model, which is well justified near the resonance, $\omega \simeq \epsilon$, and for weakly interacting system, $g \ll \omega$. Counter rotating wave term $V_2$ violates the conservation of excitation number, while conserves their parity. This term plays a central role in our treatments because it leads to the dynamical Lamb effect. Namely, non-adiabatic modulation of $V_2(t)$ provides qubit excitation with simultaneous photon creation \cite{Lozovik1,paper1}.

\begin{figure}[h]
    \center\includegraphics[width=0.95\linewidth]{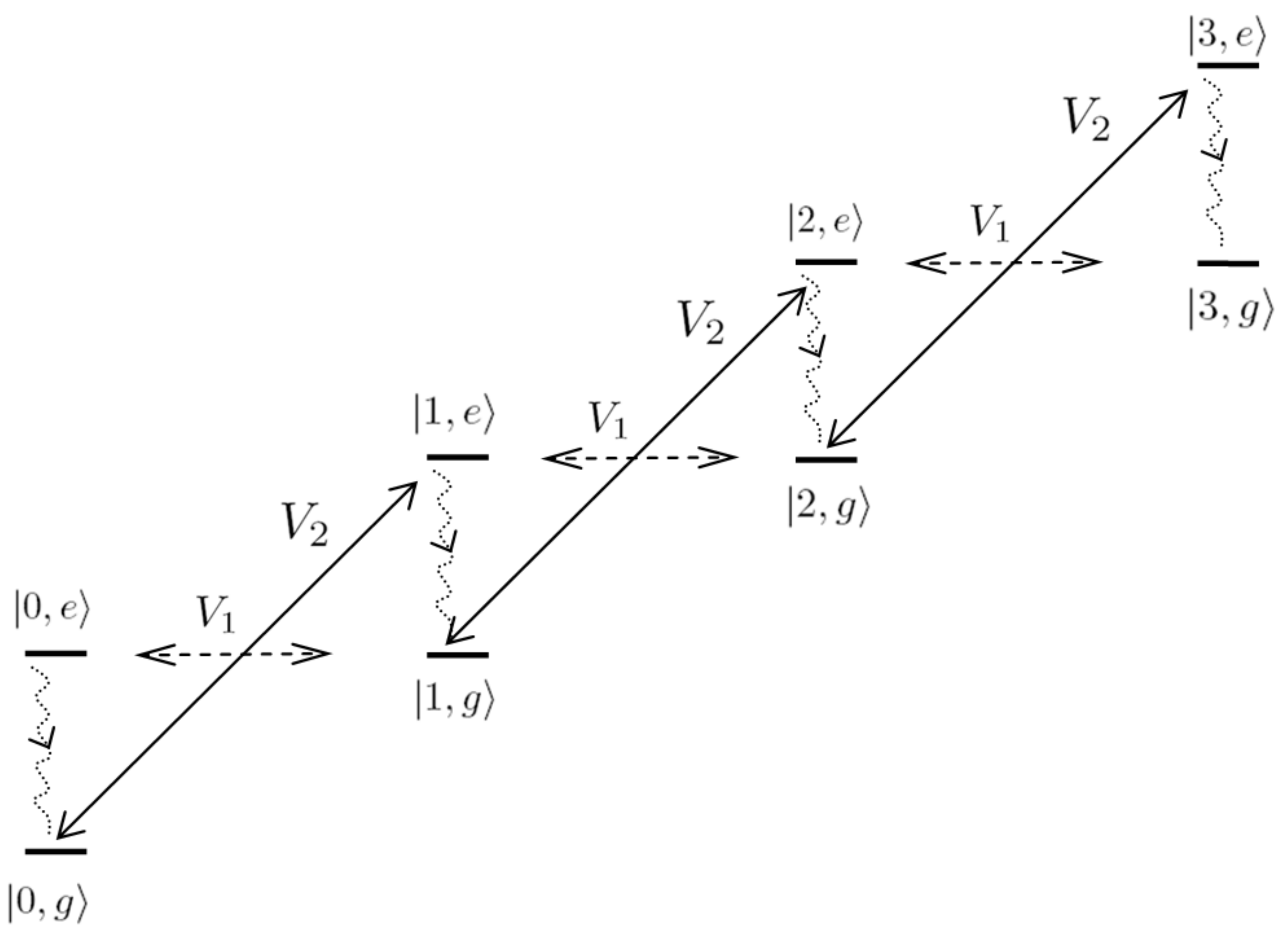}
    \caption{ The structure of bare energy spectrum in the case of a full cavity-qubit resonance and main processes induced between bare states due to interaction terms in Hamiltonian, as well as due to the decay of qubit excited state (see in the text). \label{staircase}}
\end{figure}

In our preceding paper \cite{paper1}, we took into account instantaneous
 and periodic switching of $g(t)$ of particular rectangular shapes, which provides its nonadiabatic modulation, because the dynamical Lamb effect occurs upon nonadiabatic changes of system's parameters \cite{Lozovik1}.
We considered both single switching as $g(t)=g\theta(t)$
 and periodic one as $g(t)=g\theta(\cos 2\omega t)$, where the last modulation resulted
 in a parametric pumping of the system leading to the dramatic increase of the effect in the case of a full resonance $\omega = \epsilon$. This behavior has some similarities with the so called Anti-Jaynes-Cummings regime described in Ref. \cite{ADodonov}, in which a single photon and one atomic excitation are created. In the present work, we are focused
 on the same most efficient $2\omega$-periodic modulations, whereas a particular shape of $g(t)$ can be arbitrary.
 In our solution, a major role is played by two Fourier components of the $g(t)$: $q=g_0$ and $p=g_{2\omega}$ which control the dynamics in the regime of a weak qubit-cavity coupling. For the above modulation containing periodic switching on and off, these two parameters are $p=0.5$, $q=1/\pi$. A large fraction of our results will be presented using these particular values of $p$ and $q$, since such results turn out to be rather typical for the domain of parameters $p>q$. Moreover, this choice allows us to keep a direct link with Refs. \cite{Lozovik1,paper1}. We also assume that relaxation in qubit exceeds losses in the photon mode, i.e. $\gamma\gg\kappa$,
 which might be related to an experimental situation with $\gamma\sim 1$ MHz for transmons and $\kappa\sim 10$ kHz for cavity.

The dynamics of our system in presence of energy dissipation can be found from Lindblad equation, which reads as
\begin{equation}
\partial_t\rho(t)-\Gamma(\rho(t))=-i[H(t),\rho(t)],
\label{Lindblad}
\end{equation}
where relaxation is described by means of the following matrix
\begin{eqnarray}
\Gamma(\rho)=\kappa(2a\rho a^{\dagger } - a^{\dagger }a \rho- \rho a^{\dagger }a)+\nonumber \\
\gamma(2\sigma_-\rho\sigma_+ - \sigma_+\sigma_-\rho- \rho\sigma_+\sigma_-).\label{Gamma}
\end{eqnarray}
In superconducting systems, dissipation in qubit degrees of freedom is typically much stronger than the cavity relaxation. We focus mostly on this situation, while the effect of cavity relaxation is analyzed in Section V.

In the explicit form, master equations on density matrix operator $\rho(t)$ neglecting cavity relaxation $\kappa$ read as
\begin{widetext}
\begin{eqnarray}
i \dot{\rho} _{m,n}^{gg}={\rho} _{m,n}^{gg} \omega (n-m)+i \gamma{\rho} _{m,n}^{ee}+g(t)\left(\sqrt{n}{\rho} _{m,n-1}^{ge} + \sqrt{n+1}{\rho} _{m,n+1}^{ge} - \sqrt{m}{\rho} _{m-1,n}^{eg} - \sqrt{m+1}{\rho} _{m+1,n}^{eg} \right), \nonumber \\
i \dot{\rho} _{m,n}^{ee}={\rho} _{m,n}^{ee} \left[\omega (n-m)-i \gamma\right]+g(t)\left(\sqrt{n}{\rho} _{m,n-1}^{eg} + \sqrt{n+1}{\rho} _{m,n+1}^{eg} - \sqrt{m}{\rho} _{m-1,n}^{ge} - \sqrt{m+1}{\rho} _{m+1,n}^{ge} \right), \nonumber \\
i \dot{\rho} _{m,n}^{eg}={\rho} _{m,n}^{eg} \left[\omega (n-m)- \varepsilon -i \gamma/2\right]+g(t)\left(\sqrt{n}{\rho} _{m,n-1}^{ee} + \sqrt{n+1}{\rho} _{m,n+1}^{ee} - \sqrt{m}{\rho} _{m-1,n}^{gg} - \sqrt{m+1}{\rho} _{m+1,n}^{gg} \right), \nonumber \\
i \dot{\rho} _{m,n}^{ge}={\rho} _{m,n}^{ge} \left[\omega (n-m) + \varepsilon -i \gamma/2\right]+g(t)\left(\sqrt{n}{\rho} _{m,n-1}^{gg} + \sqrt{n+1}{\rho} _{m,n+1}^{gg} - \sqrt{m}{\rho} _{m-1,n}^{ee} - \sqrt{m+1}{\rho} _{m+1,n}^{ee} \right),
\label{density-matrix}
\end{eqnarray}
\end{widetext}
where upper indices of density matrix components stand for qubit ground (g) and excited (e) states, while lower indices correspond to photon numbers. These equations can be solved numerically by truncating the basis for photon states and taking into account some reasonable number of these states. The accuracy can be verified by increasing the number of states in the basis and comparing the results with the results for a smaller basis.

However, before treating these equations, we consider a general structure of bare energy levels and processes in which they participate. These processes are due to the interaction terms $V_1$ and $V_2$ in the Hamiltonian, as well as due to the decay of the qubit excited state. Fig. 1 illustrates the dynamics of the system upon the action of the external driving in the resonant case $\omega = \varepsilon$.

In general, it may be expected that there should be a competition
between various processes in our system. Namely, there is a purely
coherent process of parametric qubit excitation tending to populate
the states $|n,e\rangle$, $n$ being odd, via the term $V_2$ of the
Hamiltonian which does not conserve the excitation number (solid
lines) and the excitation-number conserving term $V_1$ (dashed
lines). Such a process has been considered in our preceding paper
\cite{paper1}. Qubit excitation due the dynamical Lamb effect occurs
during this process thanks to $V_2$. There is also a process of a
decay of the qubit excited state (waving doted curves), which may
tend to return the system in the initial state via $V_1$. This
latter process tries to suppress the dynamical Lamb effect. However,
instead of returning to the initial state, the system can again be
excited via $V_2$ leading to the nonzero populations of the states
$|n,e\rangle$, $n$ being even. A toy model is proposed in the next
Section in order to describe some aspects of this behavior on
simplest grounds.

\section{Toy model}

Let us take into account only four bare levels, which have the
lowest energy. These are the levels $|0,g\rangle$, $|0,e\rangle$,
$|1,g\rangle$, and $|1,e\rangle$. We choose them because the system
of these four states already supports two of the most important
processes mentioned above, which are (i) an excitation of the system
via $V_2$: $|0,g\rangle \rightarrow |1,e\rangle$; (ii) a subsequent
decay $|1,e\rangle \rightarrow |0,e\rangle$ accompanied by
oscillations between $|0,e\rangle$ and $|1,g\rangle$ due to $V_1$.
It does not take into account, however, the possibility of qubit to
be excited again by $V_2$ after the decay of its excited state,
since a larger basis is needed to treat it. This process leads to
important modifications, as demonstrated in the next Section.

The system of these four levels is described by a set of 10 equations for the density matrix components. Actually, in the context of the dynamical Lamb effect, the most important components of the density matrix are those ones which are responsible for the populations of the above levels. It can be seen from the full set of equations that this set splits into two uncoupled subsets. The subset relevant for the occupation probabilities of two qubit states are
\begin{eqnarray}
i \dot{\rho} _{0,0}^{gg}=i \gamma{\rho} _{0,0}^{ee}+g(t)\left({\rho} _{0,1}^{ge}-{\rho} _{0,1}^{ge*} \right),  \label{ro1} \\
i \dot{\rho} _{0,0}^{ee}=-i \gamma{\rho} _{0,0}^{ee}+g(t)\left({\rho} _{0,1}^{eg}-{\rho} _{0,1}^{eg*} \right), \label{ro2} \\
i \dot{\rho} _{1,1}^{gg}=i \gamma{\rho} _{1,1}^{ee}+g(t)\left({\rho} _{1,0}^{ge}-{\rho} _{1,0}^{ge*} \right), \label{ro3} \\
i \dot{\rho} _{1,1}^{ee}=-i \gamma{\rho} _{1,1}^{ee}+g(t)\left({\rho} _{1,0}^{eg}-{\rho} _{1,0}^{eg*} \right), \label{ro4} \\
i \dot{\rho} _{0,1}^{eg}={\rho} _{0,1}^{eg}\left(\omega - \varepsilon - i \gamma/2 \right) +g(t)\left({\rho} _{0,0}^{ee}-{\rho} _{1,1}^{gg} \right), \label{ro5} \\
i \dot{\rho} _{0,1}^{ge}={\rho} _{0,1}^{ge}\left(\omega + \varepsilon - i \gamma/2 \right) +g(t)\left({\rho} _{0,0}^{gg}-{\rho} _{1,1}^{ee} \right).
\label{ro6}
\end{eqnarray}

We hereafter consider the qubit and cavity in a full resonance, $\omega = \varepsilon$. 
Let us represent ${\rho} _{0,1}^{ge}$ by a product of fast and slow oscillating factors as
\begin{eqnarray}
{\rho} _{0,1}^{ge}=\widetilde{{\rho}} _{0,1}^{ge} \exp (-2i\omega t),
\label{ro-split}
\end{eqnarray}
as suggested by Eq. (\ref{ro6}). All other components of the density matrix are free from fast oscillations as seen from Eqs. (\ref{ro1})-(\ref{ro5}).

Next, we insert Eq. (\ref{ro-split}) to Eqs. (\ref{ro1})-(\ref{ro6}) and perform an approximate averaging over time. For the time averaged quantities appearing in the right hand sides of resulting equations we use the following uncouplings
\begin{eqnarray}
\left\langle g(t)\widetilde{{\rho}} _{0,1}^{ge}\right\rangle_{t} \simeq \left\langle g(t)\right\rangle_{t} \widetilde{{\rho}} _{0,1}^{ge} \equiv p \widetilde{{\rho}} _{0,1}^{ge}, \nonumber \\
\left\langle g(t)\exp (-2i\omega t)\widetilde{{\rho}} _{0,1}^{ge}\right\rangle_{t} \simeq \left\langle g(t)\exp (-2i\omega t)\right\rangle_{t} \widetilde{{\rho}} _{0,1}^{ge}
\nonumber \\
\equiv  q \widetilde{{\rho}} _{0,1}^{ge},
\label{pq}
\end{eqnarray}
which are further utilized to separate fast and slow oscillations. It can be proved that they are valid provided $g(t) \ll \omega$.
Then we are going to look for a stationary, in leading order, solution that implies that the left hand sides of Eqs. (\ref{ro1})-(\ref{ro6}) must vanish. Note that we assume the time invariance of $g(t)$: $g(-t)=g(t)$.

We finally obtain the following set of linear equations for the stationary solution
\begin{eqnarray}
i \gamma{\rho} _{0,0}^{ee}+q\left(\widetilde{{\rho}} _{0,1}^{ge}-c.c. \right)=0, \\
-i \gamma{\rho} _{0,0}^{ee}+p\left({\rho} _{0,1}^{eg}-c.c. \right)=0, \\
-i \gamma{\rho} _{1,1}^{ee}+p\left({\rho} _{0,1}^{eg}-c.c. \right)=0, \\
i \gamma{\rho} _{1,1}^{ee}+q\left(\widetilde{{\rho}} _{0,1}^{ge}-c.c. \right)=0, \\
- \frac{i \gamma}{2} {\rho} _{0,1}^{eg} +p\left({\rho} _{0,0}^{ee}- {\rho} _{1,1}^{gg} \right)=0, \\
- \frac{i \gamma}{2} \widetilde{{\rho}} _{0,1}^{ge} + q \left({\rho} _{0,0}^{gg}-{\rho} _{1,1}^{ee} \right)=0.
\label{set-stationar}
\end{eqnarray}

By solving this linear system of equations, we readily express populations of the states through the population of the ground state
\begin{eqnarray}
{\rho} _{0,0}^{ee}={\rho} _{1,1}^{ee}=\frac{4q^2}{4q^2 + \gamma ^2} {\rho} _{0,0}^{gg},\\
{\rho} _{1,1}^{gg}=\frac{\gamma ^2+4p^2}{\gamma ^2+4q^2} \frac{q^2}{p^2}{\rho} _{0,0}^{gg},
\label{popul-general}
\end{eqnarray}
while ${\rho} _{0,0}^{gg}$ can be found from the normalization condition.

Let us analyze some limiting cases. We start from the case of a low dissipation, $\gamma \ll p, q$. In this case, populations of four states are all the same which is expectable. In the opposite limit, $\gamma \gg p, q$, we have:
\begin{eqnarray}
{\rho} _{0,0}^{ee}={\rho} _{1,1}^{ee} \simeq \frac{4 q^2}{\gamma ^2} {\rho} _{0,0}^{gg} \ll {\rho} _{0,0}^{gg},\\
{\rho} _{1,1}^{gg} \simeq \frac{q^2}{p^2}{\rho} _{0,0}^{gg}.
\label{strong-dissip}
\end{eqnarray}
We see that in this case the population of the qubit excited state becomes very small, while by
tuning the ratio of Fourier components $q/p$ one can redistribute the occupation probability between the state with 0 photons and 1 photon.
The higher this ratio the larger occupation of the state with 1 photon. This is a natural result in the view of the fact that $q$
is responsible for the excitation from the ground state. What is not so obvious is that ${\rho} _{1,1}^{gg}$
is dissipation-independent despite of the fact that dissipation is needed for this state to be occupied.
It is also of interest that qubit excited state does play a crucial role in such a process, nevertheless
it turns out to be essentially empty when a stationary regime is achieved. In order to achieve a ratio $q/p$ exceeding 1,
one has to use high-amplitude modulations of $g(t)$ thus changing its sign.

We also consider intermediate cases. The first one is $q \gg \gamma \gg p$. In this case, we obtain
\begin{eqnarray}
{\rho} _{0,0}^{ee}={\rho} _{1,1}^{ee} \simeq {\rho} _{0,0}^{gg},\\
{\rho} _{1,1}^{gg} \simeq \frac{\gamma ^2}{4 p^2}{\rho} _{0,0}^{gg} \gg {\rho} _{0,0}^{gg}.
\label{intermed1}
\end{eqnarray}
The first relation is expectable, since we are dealing with the strong excitation limit. However, the second one is not so trivial. It can be understood by the fact that the occupation probability is accumulated in the state $|1,g\rangle$ due to the smallness of $p$ which is responsible for the link with $|0,e\rangle$.

The second intermediate case is $p \gg \gamma \gg q$. It gives
\begin{eqnarray}
{\rho} _{0,0}^{ee}={\rho} _{1,1}^{ee} \simeq {\rho} _{1,1}^{gg} \simeq \frac{4 q^2}{\gamma^2}{\rho} \ll {\rho} _{0,0}^{gg}.
\label{intermed2}
\end{eqnarray}
This situation is rather trivial. It corresponds to the weak excitation, so that the occupation probability is accumulated in the state $|0,g\rangle$.

The toy model presented in this Section is useful since it indicates the general trend in the behavior of our system. Nevertheless, an analysis involving larger basis of bare states is certainly needed.

Note that by using an interplay between different Fourier components of the external electromagnetic signal, one can also achieve a synchronization of a qubit ensemble \cite{Rubtsov} despite of the unavoidable disorder in excitation energies of Josephson qubits \cite{Shapiro}.

\section{Full numerical analysis}

In this Section, we present the results of our numerical simulations
of the full set of equations (\ref{density-matrix}) taking into account
80 photon states. We verified the accuracy of this approximation by
increasing the number of states taken into account and comparing the
results.

\begin{figure}[h]
\center\includegraphics[width=0.95\linewidth]{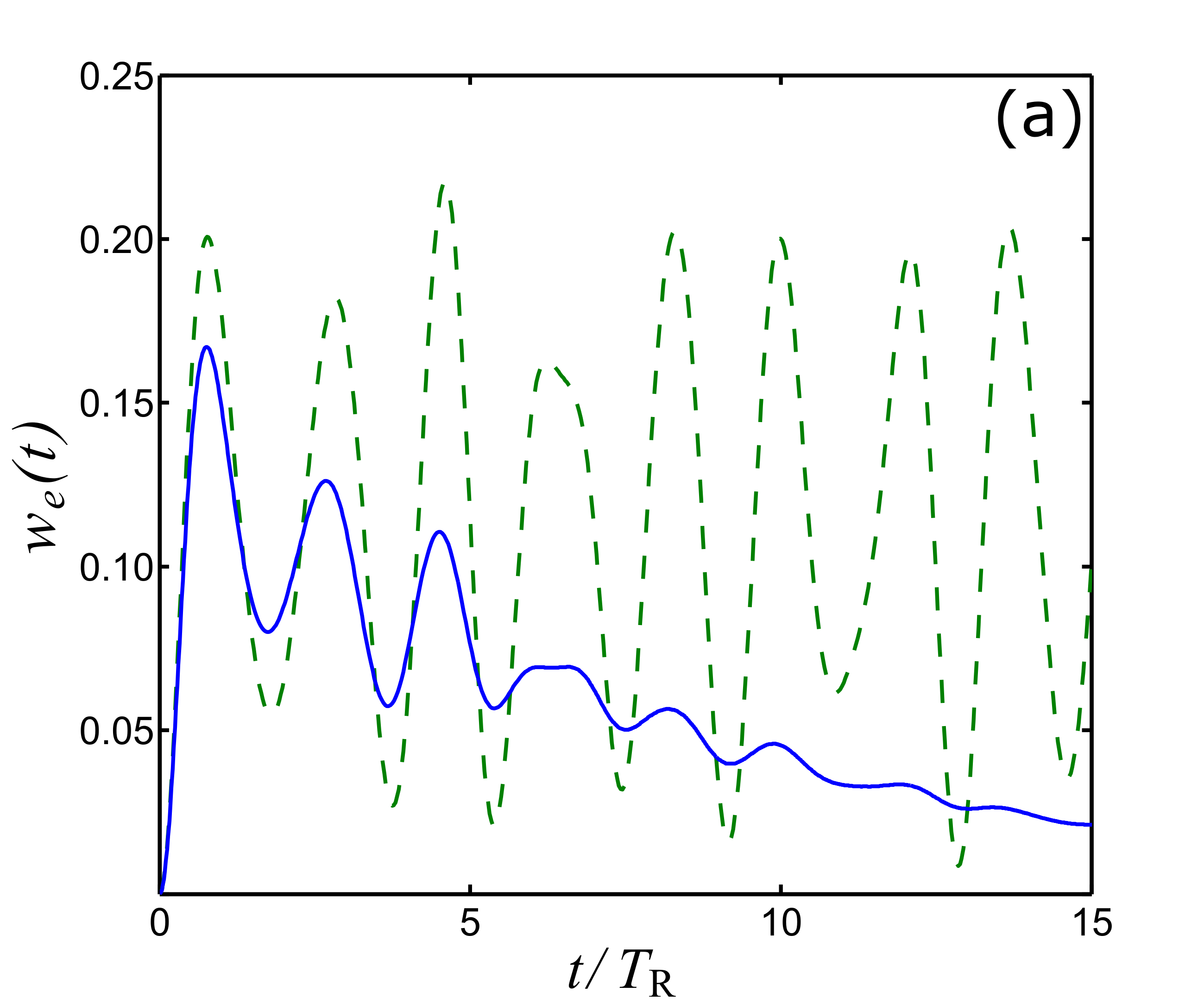}
\center\includegraphics[width=0.95\linewidth]{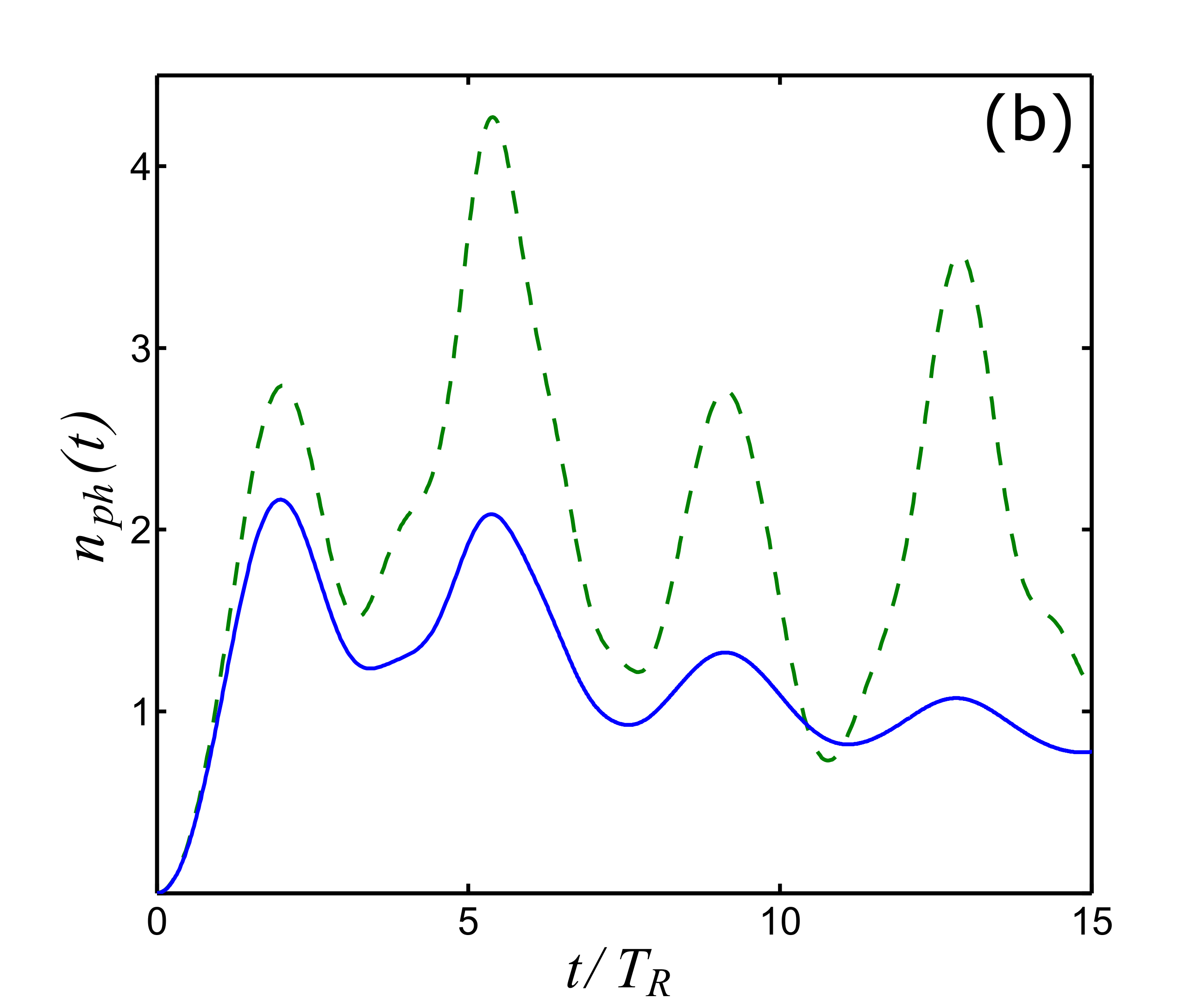} \caption{
\label{excited}(Color online) The qubit excited state population (a)
and the mean photon number (b) as functions of time after the
external parametric driving with $p=0.5$, $q=1/\pi$ is turned on at
$g_{max}=0.05 \omega $. Solid blue lines correspond to $\gamma=0.01
\omega $, while dashed green lines provide similar quantities for
$\gamma=0$.}
\end{figure}

We first consider the same modulation $g(t)=g\theta(\cos 2\omega t)$, as in our preceding paper \cite{paper1}, for which $p=0.5 < q=1/\pi$. In Fig. \ref{excited} we plot the time dependencies of the qubit excited state population $w_e$ (a) and the mean photon number $n_{ph}$ (b) after the
external driving is turned on at $g_{max}\equiv \max g(t)=0.05 \omega $, while the initial state had zero excitations, $|0,g\rangle$. The time is measured in $T_{R}=\pi/g_{max}$.
 Solid lines correspond to the case
$\gamma=0.01 \omega $, while dashed lines provide a similar quantity for
$\gamma=0$. These dependencies are actually superpositions of fast
and slow oscillations with frequencies of the order of $\omega$ and $g_{max}$,
respectively. Fast oscillations are not shown in Fig.
\ref{excited} because of their small amplitude in the limit $g_{max}/\omega \ll 1$.

It is seen from Fig. \ref{excited} that both the qubit excited state population and the mean photon number tend to experience Rabi-like
oscillations in agreement with the results of \cite{paper1}, but they
decay if nonzero $\gamma$ is taken into account. Despite of the
external driving the qubit finally saturates in its ground state.
This implies that the dynamical Lamb effect becomes suppressed at long times.
However, the mean photon number $n_{ph}$ tends to some nonzero constant value
(under the approximation neglecting losses in a photon mode).
The characteristic time of the decay of the Rabi-like oscillations
is given approximately by $1/\gamma$. The nonzero $n_{ph}(t \rightarrow \infty)$
can be treated as a residue of the dynamical Lamb effect, since the photons in the
initially empty cavity in the system we study can appear only due to the qubit excitation with simultaneous photon creation (term $V_2$) and subsequent creation of an additional photon and qubit transition to the ground state (term $V_1$).

\begin{figure}[h]
\center\includegraphics[width=0.95\linewidth]{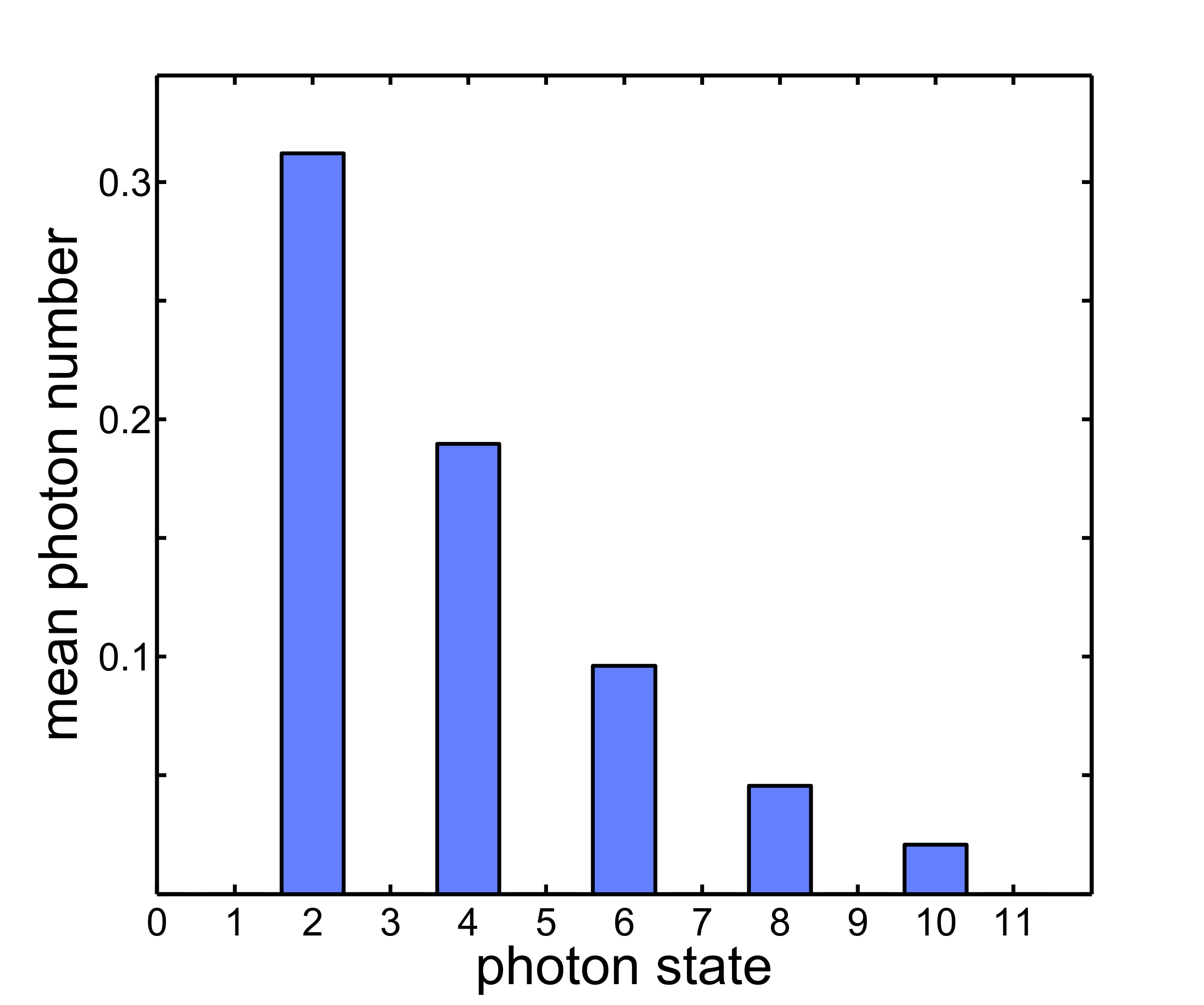} \caption{
\label{photon-distr}(Color online) A histogram for the mean number
of photons in the $n$-photon states after the stabilization at
$\gamma=0.01 \omega $, $g_{max}=0.05 \omega $. The parameters of a
modulation of $g(t)$ are $p=0.5$, $q=1/\pi$.}
\end{figure}

\begin{figure}[h]
\center\includegraphics[width=0.95\linewidth]{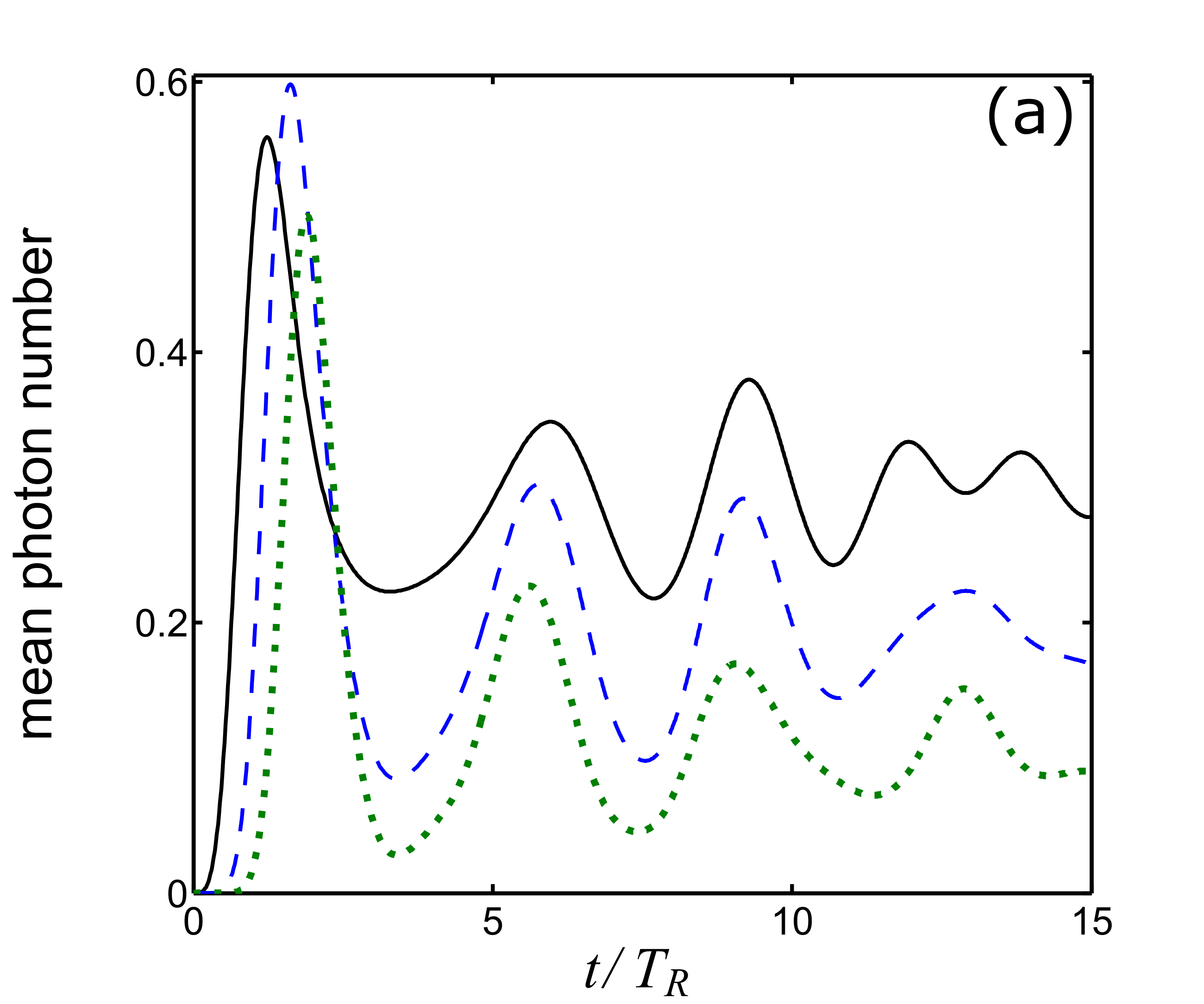}
\center\includegraphics[width=0.95\linewidth]{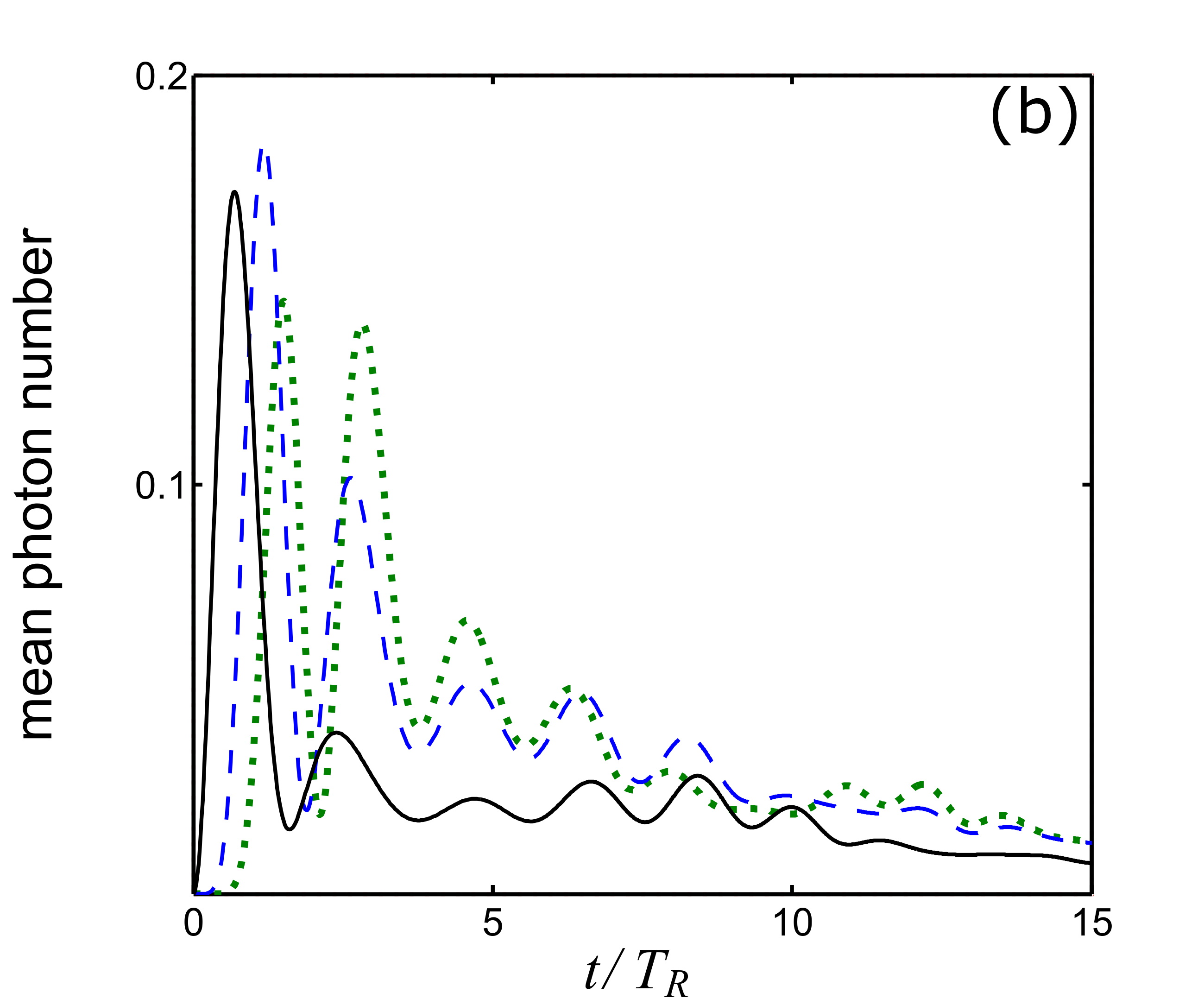} \caption{
\label{dyn-phot}(Color online) Dynamics of mean photon numbers in
$n$-photon  states with even (a) and odd (b) values of $n$ at
$\gamma=0.01 \omega $, $g_{max}=0.05 \omega $. The parameters of the
modulation of $g(t)$ are $p=0.5$, $q=1/\pi$. Solid black, dashed
blue, and dotted green lines in (a) correspond to 2-, 4-, and
6-photon states, respectively, while similar lines in (b) correspond
to 1-, 3-, and 5-photon states.}
\end{figure}

The statistics of photons states
after their stabilization turns out to be rather peculiar and it is
definitely dictated by a parametric excitation of photons. In Fig.
\ref{photon-distr} we plot a histogram for the dependence of the
mean number of photons in the $n$-photon state on $n$. We see that
only the states with even values of $n$ are populated after the
stabilization. Fig. \ref{dyn-phot} shows how photon states become
stabilized after the parametric driving is turned on. The set of parameters
for these two figures is the same as in the case of Fig. \ref{excited}.

Figs. \ref{excited} and \ref{photon-distr} evidence that only
low-energy photon states and ground state energy of a qubit are
populated after the stabilization, i.e., the states $|n,g\rangle$
with $n\sim 1$. This is due to the competition between two
processes, as can be seen already from the simple toy model
presented in the preceding Section. The fact that we do have a
stabilization in our system with qubit being in its ground state
means that the process, involving decay, is stronger. Nevertheless,
in order to achieve such a stabilization, as shown in Figs.
\ref{excited} and \ref{dyn-phot}, certain intermediate dynamical
regime is needed for which qubit can be in its excited state.

There can be seen a certain analogy between this final regime  and a phenomenon of a parametric down conversion. In both cases, an external pump of the system by a periodic signal results in a spontaneous generation of pairs of photons with lower frequencies. As known, microscopic mechanisms underlying such nonlinear effects can be different, see, e.g., Ref. \cite{Boyd}. In this paper, we are mostly interested in such a microscopic description of a particular system suitable for the realization of the dynamical Lamb effect. Indeed, the main focus of our work is on qubit degrees of freedom, while in the theory of a parametric down conversion the main emphasis is on a photon generation, whereas atomic degrees of freedom are normally considered as a source of nonlinearities. Apart of the interest from the viewpoint of fundamental physics, our approach is also motivated by a perspective to use such systems in quantum technologies, in which qubit degrees of freedom as well as correlations between them and photon modes are of crucial importance.

The results obtained by numerical simulations are in a qualitative
agreement with the results of our toy model in the strong $\gamma$
limit at $p > q$. Namely, we see that qubit excited state tends to become
empty, while photon states with lower energy have larger
populations. Of course, the toy model is unable to correctly
describe other important features because of the very strong
truncation of the basis encoded in it. For instance, within our toy model, both even and odd photon states are populated in the final state. In order to see how the depletion of odd states emerges as basis size increases, we extend this basis step by step by taking into account more and more levels and solving the problem numerically. We then follow the evolution of level populations in the final state.

By considering six levels, we see that the population of the level $|1,e\rangle$ decreases. The reason is that this occupation probability is redistributed from this level to the "new" state $|2,g\rangle$ through $V_1$, while the only channel to increase it is an excitation via $V_2$ from $|1,g\rangle$ to $|2,e\rangle$ with a subsequent decay of the qubit excited state. Since we are in a regime when $V_1$ overcomes $V_2$, the first process dominates and leads to a partial depletion of $|1,e\rangle$. We then see that, due to this mechanism, the larger basis we take into account the stronger the effect of depletion of the states with qubit excited. Infinite basis leads to essentially full depletion of these levels. However, in this case, only states with even photon numbers and qubit in the ground state can be populated, since there is a certain asymmetry between the two subsets of levels with even and odd numbers of photons. Indeed, the "ladder" of states in Fig. 1 starts with zero (even) number of photons. The state $|0,e\rangle$ can be empty only if $|1,g\rangle$ is empty. Then, we can repeat the same argument to $|2,e\rangle$ and $|3,g\rangle$ etc. to see that only populations of states with even photon numbers do survive after the stabilization.

Note that only even states are also populated under the action of the dynamical Casimir effect.
Thus, if generation of Casimir photons is not completely ruled out during an experiment due to some drawbacks of experimental setup, this fact makes it not easy
to distinguish between the dynamical Lamb effect and dynamical Casimir effect via photons,
provided photon states are studied in experiments at $t \gtrsim 1/\gamma$.
Hence, one has to perform measurements within the time interval $\lesssim 1/\gamma$, when both even and odd photon states are populated (as well as qubit excited state), in contrast to the photon statistics due to the dynamical Casimir effect.

\begin{figure}[h]
\center\includegraphics[width=0.95\linewidth]{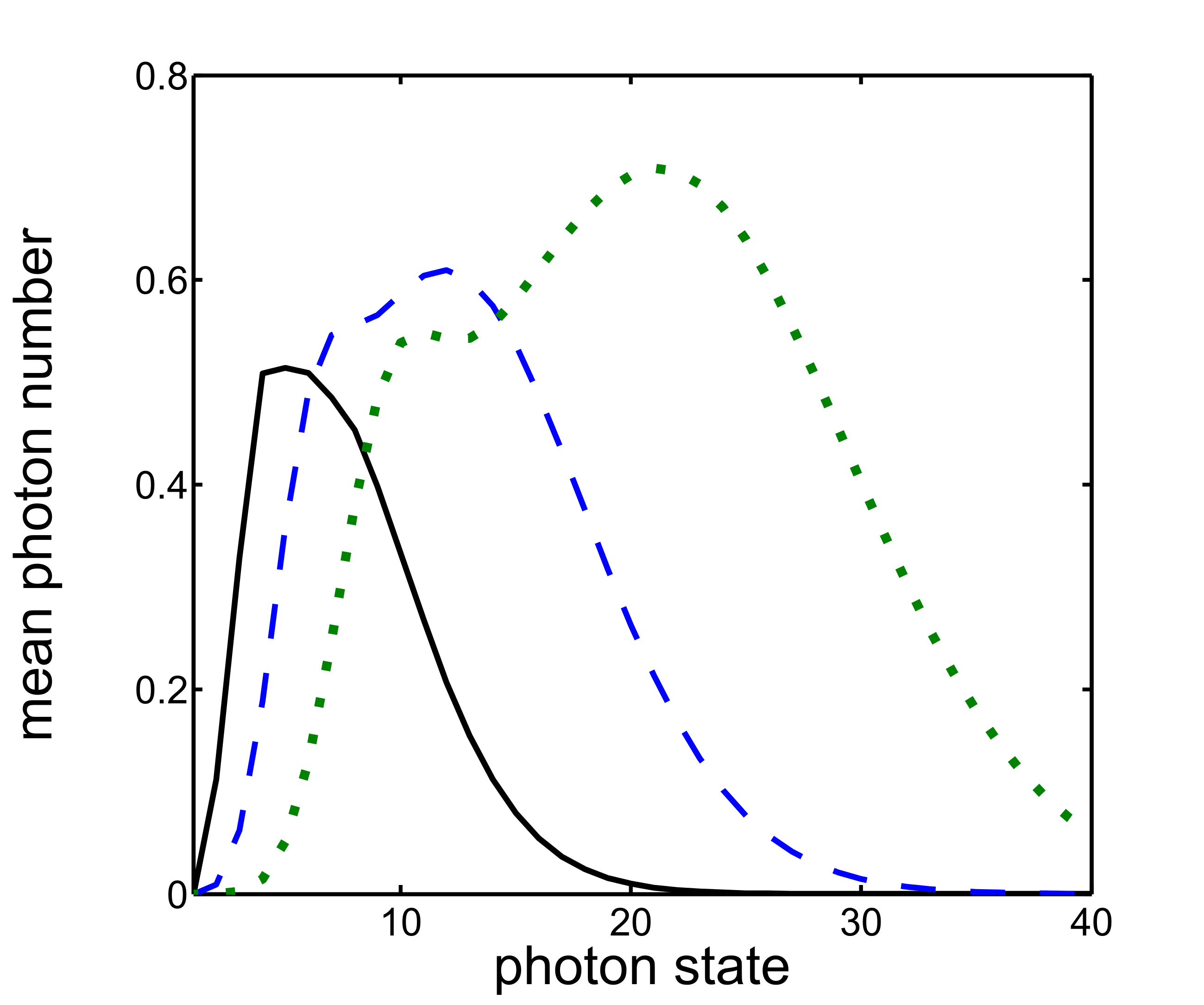} \caption{
\label{run}(Color online) The dependence of the mean photon in the
$n$-photon state number as a function of $n$ in three different
moments of time after the switching on the parametric driving at
$\gamma=0.01 \omega $, $g_{max}=0.05 \omega $. The parameters of
modulation of $g(t)$ are  $p=0.3$, $q=1$. Solid black, dashed blue,
and dotted green lines correspond to the moments of time $T_{R}$,
$5T_{R}$, and 10 $T_{R}$, respectively.}
\end{figure}

Nevertheless, there is a
method to make a dynamical Lamb effect much more pronounced at $t \gtrsim 1/\gamma$.
We can see that our toy model predicts some change of the behavior
for  those types of the external drive, for which $q$ exceeds $p$, which implies a usage
of sign-alternating time dependences of coupling constant.
We now examine this threshold in numerical simulations. In Fig.
\ref{run} we plot the mean photon number in the $n$-photon state as
a function of $n$ in three different moments of time after the
switching of parametric driving at $p=0.3$, $q=1$. We see no
stabilization at this ratio of $p/q$. Namely the maximum of this
dependence increases with time, so that the total mean photon number
also grows. This feature can be again traced from our toy model
which predicts, in the strong $\gamma$ limit, certain change of the
behavior. The difference is that the toy model includes only four
states, so that there is a boundary above which the maximum cannot
move. Within the toy model, we see a tendency to maximize the
probability of finding a system in the one-photon state, while in
reality this maximum starts to move further. Again, this change of
the behavior at $p=q$ can be tested in experiments.

Let us mention that, in the case of a composite system, a possibility to make quantum effects quite robust by using a periodic pumping of a coupling constant between its constituent parts was demonstrated in Ref. \cite{Galve} for two coupled harmonic oscillators at finite and high temperatures. In contrast, we consider a coupled system consisting of two parts which obey Bose and Pauli statistics, respectively, and at low temperatures. Nevertheless, our results together with the results of Ref. \cite{Galve} indicate that we here deal with a certain class of related phenomena.

\begin{figure}[h]
\center\includegraphics[width=0.95\linewidth]{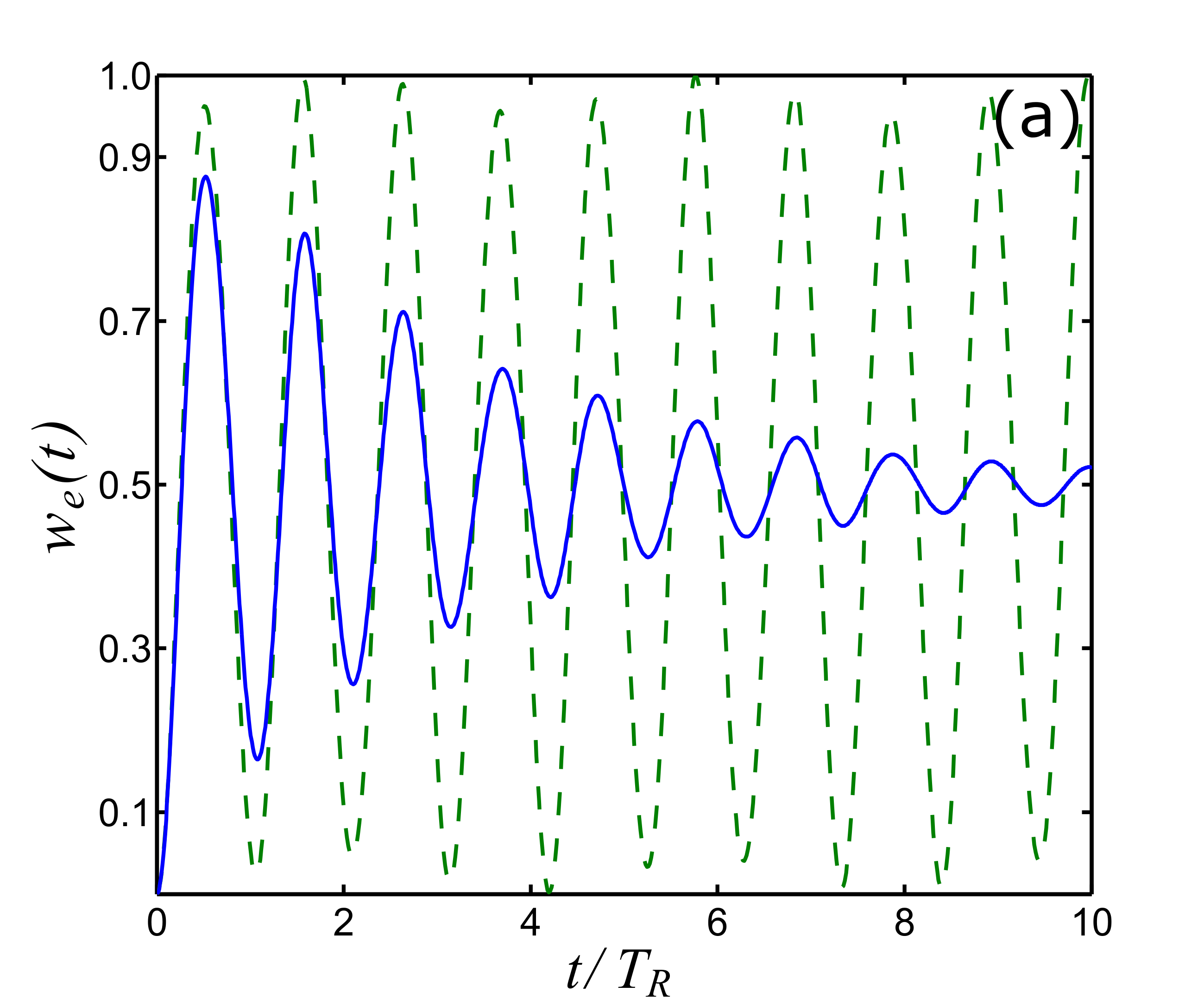}
\center\includegraphics[width=0.95\linewidth]{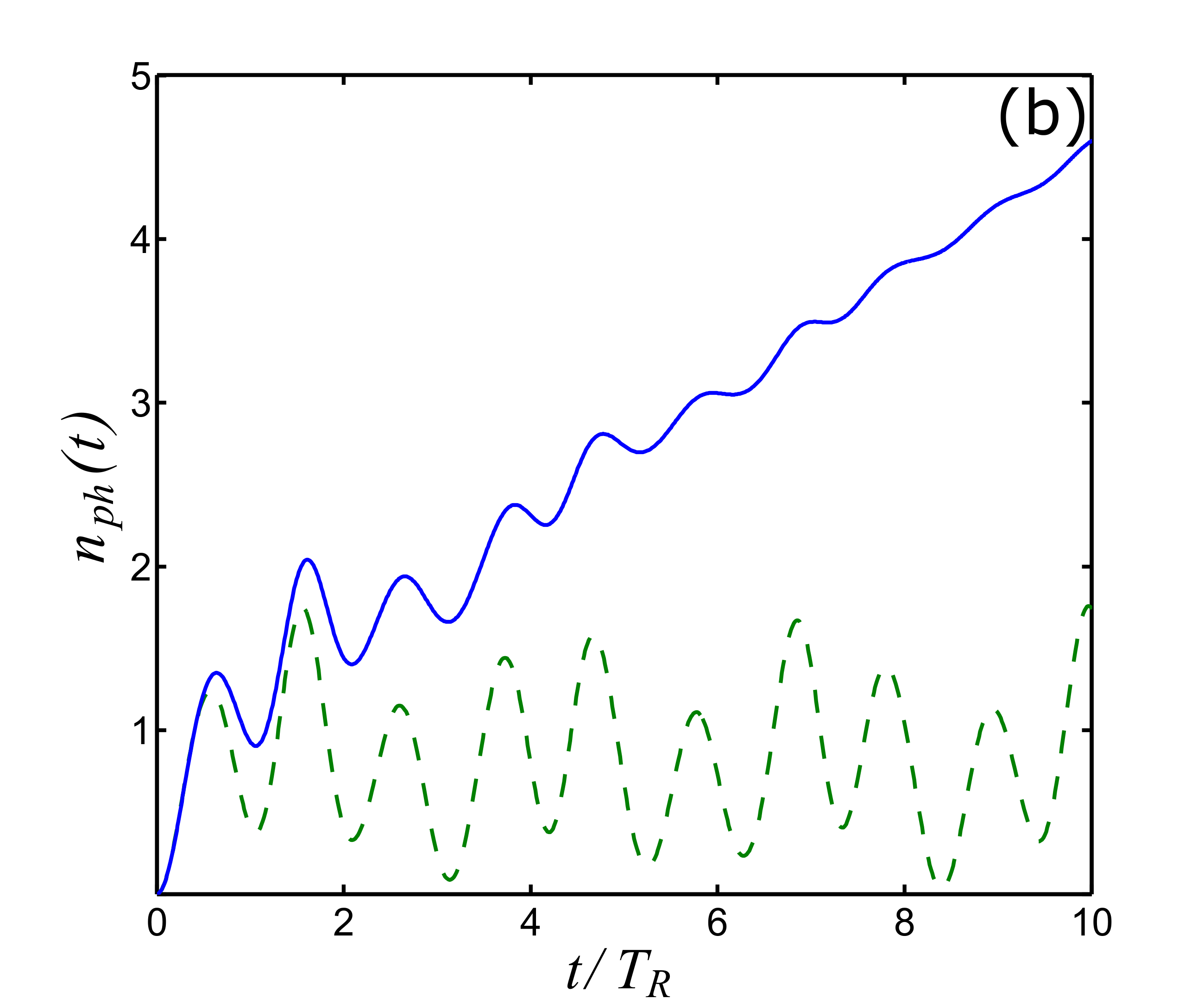} \caption{
\label{run1}(Color online) The dependence of the qubit excited state
population  (a) and mean photon number (b) on time after the
switching on the parametric driving at $g_{max}=0.05 \omega $,
$p=0.3$, $q=1$ at nonzero dissipation $\gamma=0.01 \omega $ (solid
blue lines) and $\gamma=0$ (dotted green line).}
\end{figure}

Fig. \ref{run1} shows the time evolution of the qubit excited state population (a) and mean photon number (b) after the turning on the
parametric driving characterized by parameters $p=0.3$, $q=1$ and at $g_{max}=0.05 \omega $.
Solid lines correspond to the
dissipative system with $\gamma=0.01 \omega $, while dashed curves provide
similar dependencies for $\gamma=0$. We see that finite value of
$\gamma$ leads to the decay of the Rabi-like oscillations for the
qubit excited state population which tends to be stabilized at the value
$1/2$ and does not vanish. Thus, the dynamical Lamb effect is much more robust with respect to
the dissipation in this case. The most striking feature is that after some initial oscillations, the mean photon number
starts to grow linearly which shows that there is no
stabilization at this type of parametric driving. It is remarkable that this counterintuitive growth is
only possible if energy dissipation due to the decay of the qubit excited state is present in the system. Indeed, such a growth is absent if $\gamma=0$, as seen from Fig. \ref{run1}. This happens because a new channel of photon generation is open provided $\gamma$ is finite, see Fig. 1. Such a channel does not exists within our toy model because of the basis truncation. It consist in excitation of the initial configuration via $V_2$, the subsequent decay of the qubit excited state, and again in excitation of the qubit from the ground to the excited state with simultaneous photon creation. Hence, both even- and odd-number photon states become populated. Thus, because of the strong increase of the mean photon number, a dissipation-assisted parametric amplification of vacuum occurs.

Let us discuss in a more detail a crossover between the two types of behavior which
occurs at $p=q$, as deduced from numerics. We continue the set of equations (\ref{ro1})-(\ref{ro6}) by taking into account all photon states. We also know that for
$\xi=q/p<1$ at long time $t\gg 1/\gamma$ only
$\rho^{gg}_{2n,2n}$ is nonzero, while all remaining relevant components vanish.
We then equate these quantities
to zero in the new set of equations, as well as time derivatives $\partial
\rho / \partial t$. This leads to following recurrent
relations
\begin{eqnarray}
\rho_{m+1,n}^{gg}=-\xi\sqrt{\frac{m}{m+1}}\rho_{m-1,n}^{gg}, \\
\rho_{m,n+1}^{gg}=-\xi\sqrt{\frac{n}{n+1}}\rho_{m,n-1}^{gg}.
\end{eqnarray}

These equations readily yield the identity:
\begin{eqnarray}
\rho_{n+2,n+2}^{gg}=\xi^2 \frac{n+1}{n+2}\rho_{n,n}^{gg}.
\end{eqnarray}

Starting from $\rho_{0,0}^{gg}$, we obtain:
\begin{eqnarray}
\rho_{2j,2j}^{gg}=\xi^{2j}\frac{(2j-1)!!}{(2j)!!}\rho_{0,0}^{gg},
\end{eqnarray}
$j=1,2, \ldots \infty$. This recurrent relation is in a full agreement with our results of numerical simulations for density matrix diagonal components.

In order to determine $\rho_{0,0}^{gg}$, we use a normalization
condition $\mathop{\mathrm{Sp}}\nolimits \rho \equiv 1$, which can be rewritten as
\begin{eqnarray}
\left( 1+\sum_{j=1}^{\infty} \xi^{2j}\frac{(2j-1)!!}{(2j)!!} \right)
\rho_{0,0}^{gg}\equiv 1.
\end{eqnarray}
This series converges provided $\xi<1$ or equivalently $q<p$, otherwise the normalization
is impossible. In the case $\xi<1$, the result of the summation is
\begin{eqnarray}
\left(
1+\frac{\xi^2}{\sqrt{1-\xi^2}\left(1+\sqrt{1-\xi^2} \right)}
\right)
\rho_{0,0}^{g,g}\equiv 1,
\end{eqnarray}
as can be directly checked by performing a Taylor expansion.

Thus, the stationary solution indeed
exists if the condition $\xi<1$ is satisfied, while there is no
stationary solution at $q \ge p$.

\section{Effect of cavity relaxation}

In this section, we take into account cavity dissipation, which is typically much smaller than dissipation in qubit degrees of freedom in available superconducting qubit-cavity systems. Nevertheless, its effect can be of importance in the view of the fact that different types of dissipation may open various channels in the dynamics of the system, as we have seen in the preceding sections. Thus, we take into account nonzero $\kappa$ in the Lindblad equation, as given by Eqs. (\ref{Lindblad}), (\ref{Gamma}).

\begin{figure}[h]
\center\includegraphics[width=0.95\linewidth]{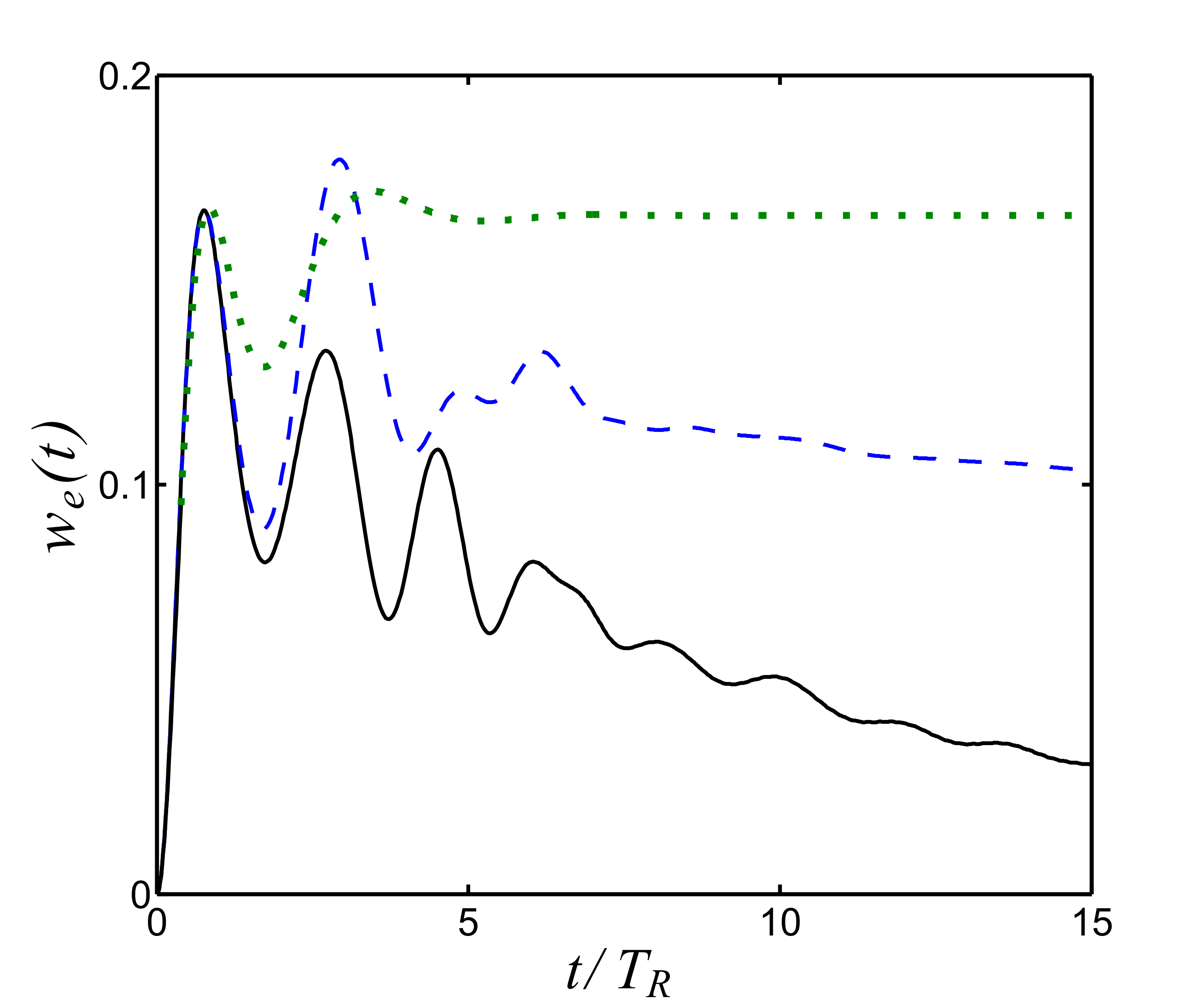} \caption{
\label{excited-kappa}(Color online) The qubit excited state
population at $p=0.5$, $q=1/\pi$, $g_{max}=0.05 \omega $,
$\gamma=0.01 \omega $ and at three different values of $\kappa /
\gamma$, which are 0.01 (solid black line), 0.1 (dashed blue line),
and 1 (dotted green line).}
\end{figure}

We start with the consideration of the modulation with $q < p$, for
which the qubit excited state population vanishes at $\kappa = 0$
and at $t \rightarrow \infty$. Fig. \ref{excited-kappa} shows this
quantity as a function of time at $p=0.5$, $q=1/\pi$, $g_{max}=0.05
\omega $, $\gamma=0.01 \omega $ and at three different values of
$\kappa / \gamma$, which are 0.01 (solid black line), 0.1 (dashed
blue line), and 1 (dotted green line). Remarkably, finite cavity
dissipation leads to the nonzero qubit excited state population at
long time. This happens because cavity relaxation tends to decrease
the mean photon number without changing the state of the qubit.
Therefore, if the qubit is in excited state, instead of its
relaxation to the ground state controlled by $\gamma$, the state of
the whole system can be changed by decreasing the photon number and
keeping qubit excited. In other words, there is a certain
competition between $\gamma$ and $\kappa$ in this case. Hence,
higher cavity dissipation can also help to increase the qubit
excited state population at $t \rightarrow \infty$ thus supporting
the dynamical Lamb effect.

\begin{figure}[h]
\center\includegraphics[width=0.95\linewidth]{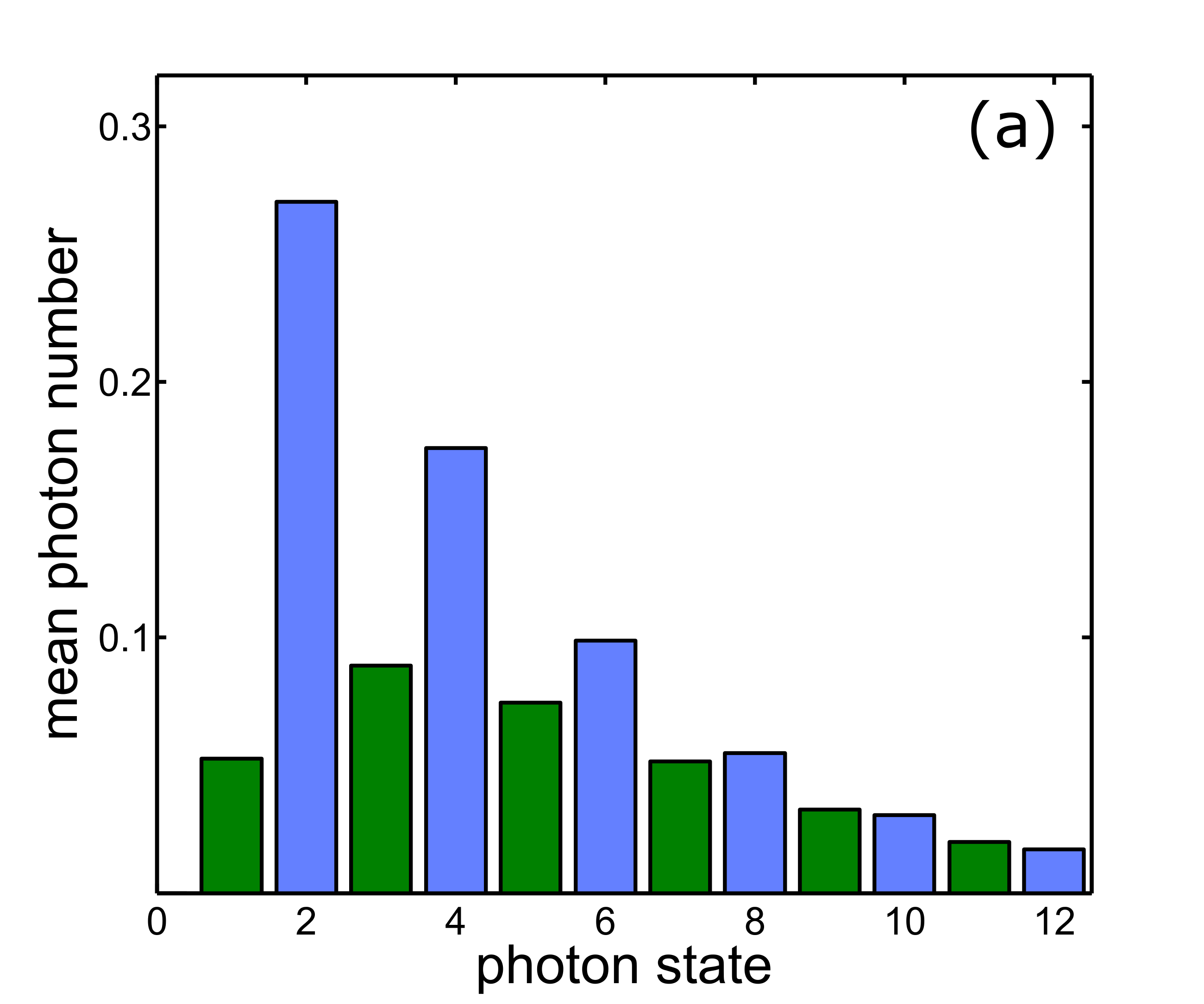}
\center\includegraphics[width=0.95\linewidth]{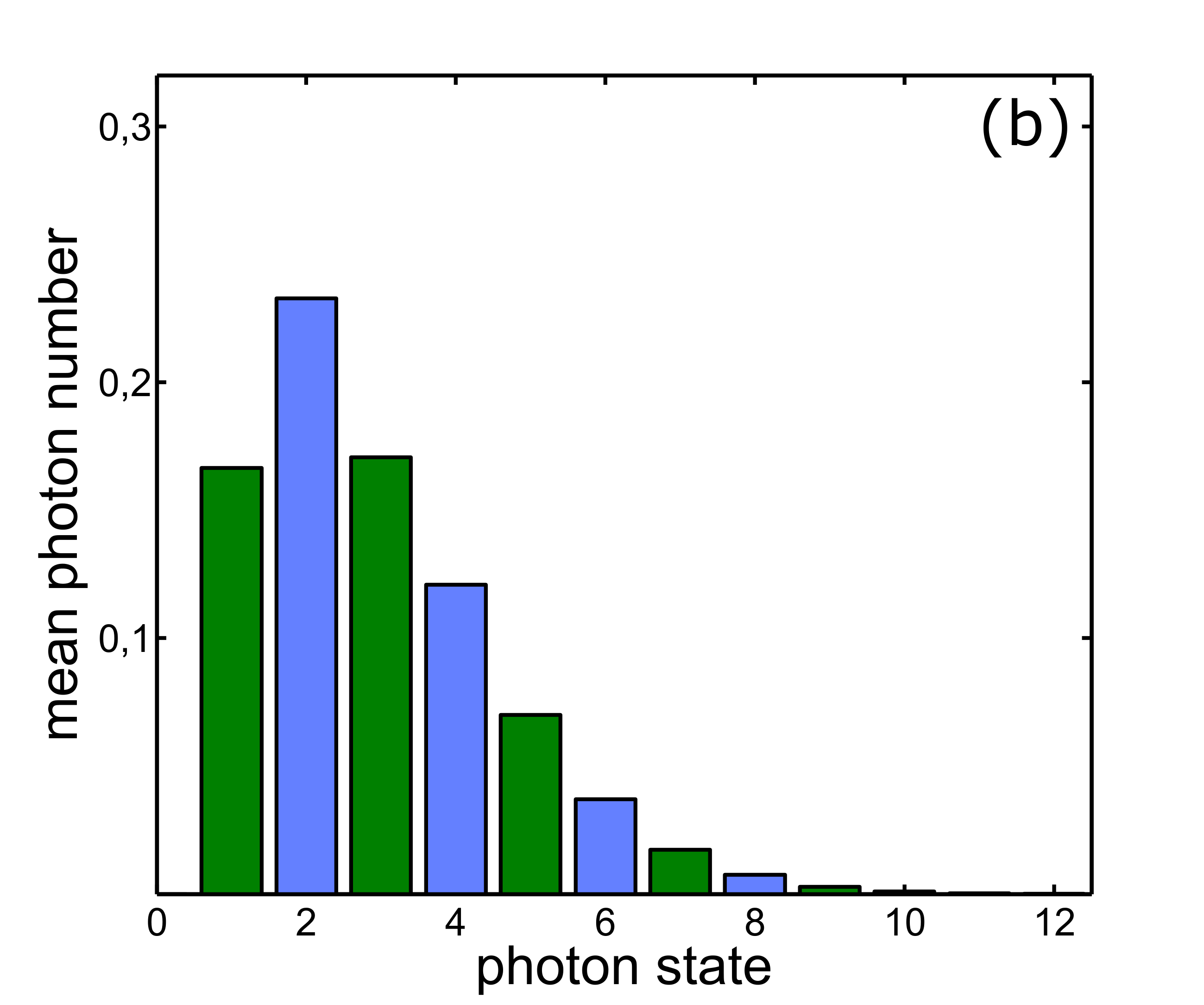} \caption{
\label{photon-distr-kappa}(Color online) Histograms for the mean
number of photons in the $n$-photon states after the stabilization
at $\gamma=0.01 \omega $, $g_{max}=0.05 \omega $, and at two
different values of $\kappa / \gamma$, which are 0.1 (a) and 1 (b).
The parameters of a modulation of $g(t)$ are $p=0.5$, $q=1/\pi$.
Dark grey (green) bars correspond to odd $n$, while light grey
(blue) bars refer to even $n$.}
\end{figure}

Fig. \ref{photon-distr-kappa} shows histograms for the mean number of photons in the $n$-photon states after the stabilization at $\gamma=0.01 \omega $, $g_{max}=0.05 \omega $, $p=0.5$, $q=1/\pi$ and at two different values of $\kappa / \gamma$, which are 0.1 (a) and 1 (b). We see that states with odd values of $n$ start to be populated because of the processes, which change photon number without changing qubit state. However, populations of these state remain small at $\kappa / \gamma \ll 1$.

\begin{figure}[h]
\center\includegraphics[width=0.95\linewidth]{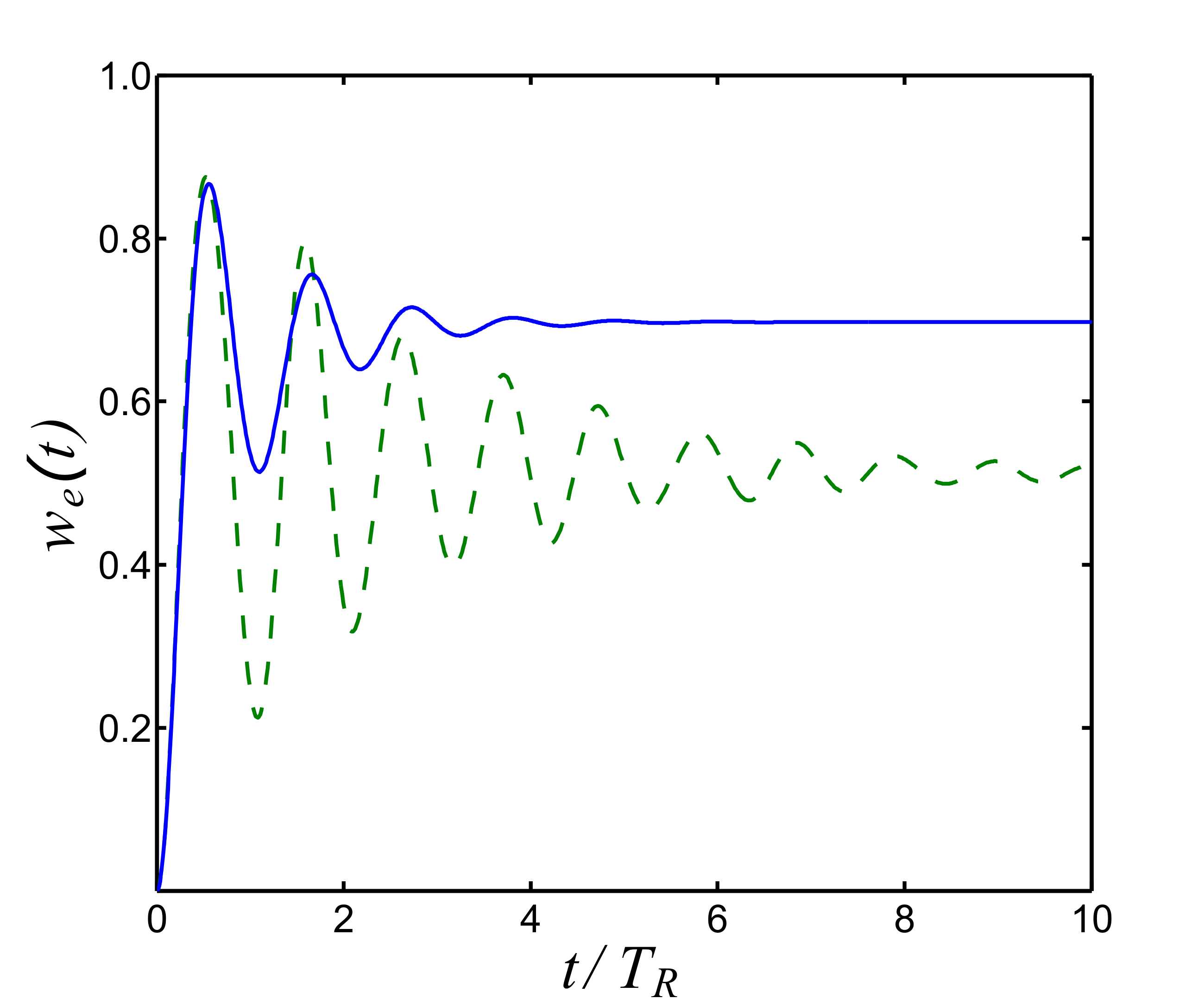} \caption{
\label{excited-kappa-strong}(Color online) The qubit excited state
population at $p=0.3$,  $q=1$, $g_{max}=0.05 \omega $, $\gamma=0.01
\omega $ and at two different values of $\kappa / \gamma$, which are
1 (solid blue line) and 0.1 (dashed green line).}
\end{figure}

\begin{figure}[h]
\center\includegraphics[width=0.95\linewidth]{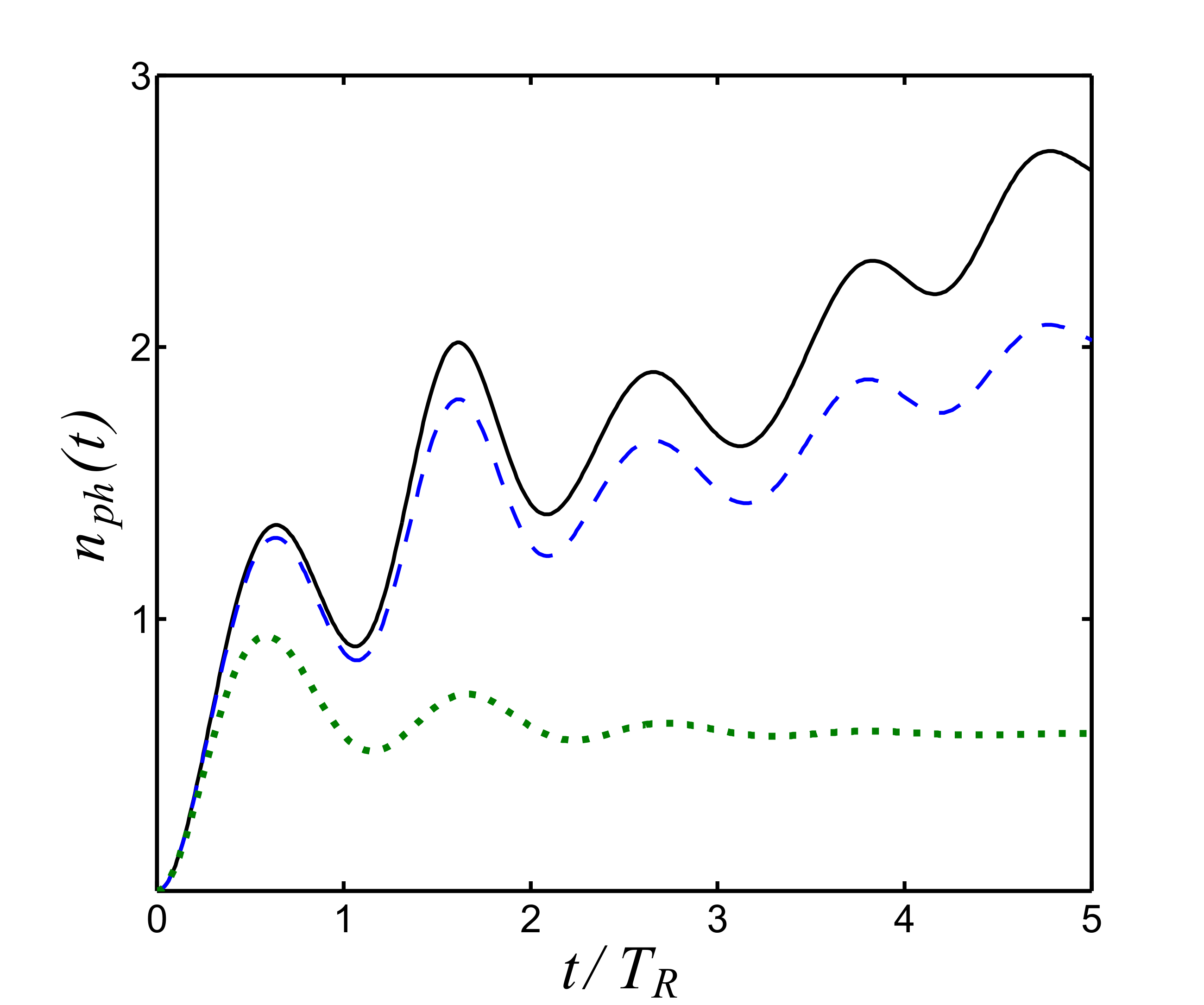} \caption{
\label{excited-kappa-strong-photon}(Color online) Mean photon number
at $p=0.3$, $q=1$, $g_{max}=0.05 \omega $, $\gamma=0.01 \omega $ and
at three different values of $\kappa / \gamma$, which are 0.01
(black solid line), 0.1 (blue dashed line), and 1 (green dotted
line).}
\end{figure}

We now consider modulations with $q > p$. Fig. \ref{excited-kappa-strong} shows $w_e (t)$ at two different values of $\kappa$. We again see that $w_e (t \rightarrow \infty)$ grows, as $\kappa$ increases, but this growth is rather weak at $\kappa / \gamma \ll 1$. This is also due to the processes which change photon number without affecting qubit degrees of freedom. We again arrive at the same conclusion, as in the case $q < p$ that cavity dissipation increase the dynamical Lamb effect within our scheme at $t \rightarrow \infty$. Actually, these two different cases, $q > p$ and $q < p$, become not so distinct when nonzero $\kappa$ is taken into account, as can be expected. Indeed, Fig. \ref{excited-kappa-strong-photon} shows mean photon number as a function of time for three different values of $\kappa$. The nearly linear growth of this quantity at $t \rightarrow \infty$ found for $\kappa=0$ is replaced by its saturation. Its final value drops, as $\kappa$ grows. However, it still can be much larger than the same quantity in absence of any dissipation, which implies that an additional channel of photon generation from vacuum with assistance of qubit relaxation, as discussed in the preceding Section, still exists in this $\kappa \neq 0$ case.

The major result of this Section is that nonzero cavity dissipation increases the qubit excited state population at $t \rightarrow \infty$. This feature can be used as an alternative tool to increase the effect without switching to sign-alternating modulations which can be not easy to implement in experiments.

Let us mention that some aspects of the temporary evolution of systems with dynamically tunable light-matter interaction were already studied in literature, see, e.g., Refs. \cite{Liberato, Garziano, Benenti, Brandes, VDodonov, ADodonov, ADodonov1}. This interaction was assumed to be either variated along or simultaneously with other parameters, such as a cavity frequency. In most of these studies, however, only a regime of weak modulation was considered. In general, it was also implicitly suggested that dissipation rates of both qubit and photon states are of the same order. Due to these assumptions, a rich dynamical picture, predicted in the present paper, has not been revealed up to now, to the best of our knowledge. Moreover, as usual in the case of nonlinear optical effects, these studies were mainly focused on the analysis of properties of photons generated upon the modulation of system parameters. For instance, in Ref. \cite{ADodonov1} the interaction between the Casimir photons and matter was shown to be responsible for a nonlinear in photon number term of purely photonic effective Hamiltonian in certain limits. The latter was obtained by exclusion of atomic degrees of freedom from the full "microscopic" system Hamiltonian. Within this approach, a term "nonlinear dynamical Casimir effect" was introduced in Ref. \cite{ADodonov1}. In contrast, the present paper as well as preceding articles \cite{Lozlett,Lozovik1,paper1} are concentrated mostly on what goes on with qubit (atom) degrees of freedom. From this perspective, the nonlinear dynamical Casimir effect in a nonstationary cavity is intrinsically related with the atom excitation due to absorption of Casimir photons as well as with the dynamical Lamb effect.

\section{Numerical solution for steady state limit}

In this Section, we provide an alternative approach to the problem, which enables us to directly attain the steady state limit achieved after the stabilization of the system and to crosscheck our results.

Solution for the density matrix elements in the steady state limit, i.e. on time scales exceeding significantly  relaxation times, can be found numerically without direct integration  over  the entire evolution period.  This calculation can be performed by means of an integration  of the Lindblad equation over a single period of  time-dependent  Hamiltonian $H(t)$, i.e. within the interval $0<t<\pi/ \omega$. In this solution we do not perform integrating out of fast oscillating terms and do not use transition to rotating frames.

The Lindblad  equation can be rewritten through the supermatrix $A(t)$ acting on vector $\vect{\rho}(t)$ combined from elements of the density  matrix $\rho(t)$
\begin{equation}
d\vect{\rho}(t)/dt=A(t)\vect{\rho}(t).
\end{equation}
In the steady state limit we assume that this solution is periodic $\vect{\rho}(t+\pi/\omega)=\vect{\rho}(t)$ with the period  $T=\pi/\omega$ of Hamiltonian $H(t)$ and $A(t)$. We find numerically matrix of evolution $U$ which relates $\vect{\rho}(T)$ and $\vect{\rho}(0)$
\begin{equation}
\vect{\rho}(T)=U\vect{\rho}(0).
\end{equation}
Eigenvector of the $\vect{\rho}_0=U\vect{\rho}_0$ gives steady state solution $\vect{\rho}_0=\vect{\rho}(NT)$ realized at infinite limit of $N$. Integration of the Lindblad equation over $0<t<\pi/ \omega$ with the initial condition $\vect{\rho}_0$ provides periodic steady solution   $\rho_{st}(t)$. Averaging of the diagonal elements of $\rho_{st}(t)$ over the time provides levels populations in qubit and photon channels. This solution gives  mean  photon numbers, which are fully identical to the above results of a time-dependent numerical solution.

\section{Summary and conclusions}
 \label{summary}

A coupled system of a superconducting qubit and microwave resonator can be used for experimental observation of the dynamical Lamb effect \cite{paper1} which can be treated as a parametric excitation of an atom due to the nonadiabatical modulation of its Lamb shift \cite{Lozovik1}. This can be achieved by dynamically tuning the vacuum Rabi frequency (the strength of the coupling between the qubit and resonator) without changing all other parameters, such as a resonator frequency. Under these conditions, no generation of Casimir photons takes place, which is a crucial condition for the isolation of the dynamical Lamb effect from other nonstationary QED phenomena also leading to the parametric excitation of a qubit. Such a modulation of vacuum Rabi frequency in superconducting circuits is possible thanks to several approaches proposed recently \cite{tunable1,tunable2,tunable3}. Notice that in contrast to natural systems, it is also possible to achieve a regime of strong or even ultra-strong light-matter coupling in artificial superconducting systems.

In the present paper we studied an influence of energy dissipation on qubit excitation due to the dynamical Lamb effect. An influence of dissipation in a qubit is of particular importance since it leads to qubit de-excitation and it also far exceeds cavity relaxation in typical superconducting qubit-cavity systems.

Our major conclusion is that the qubit excited state population in presence of dissipation depends crucially on the character of the vacuum Rabi frequency modulation. Note that we assumed that qubit and resonator in the initial moment were not excited. We also took into account that a decay of photon states in superconducting circuits is typically much weaker than relaxation in a qubit which allows for the separation of characteristic time scales of two types of dissipation.

We found that some types of periodic modulation of the vacuum Rabi frequency lead to the decay with time of the qubit excited state population, while the mean number of generated photons tends to be stabilized around some finite number. However, other types of parametric driving of the same quantity lead to a completely different behavior. In this case, the qubit excited state population state becomes stabilized near the large value of $1/2$, while the number of photons in the system grows nearly linearly with time until it also becomes stabilized by photon field relaxation. Hence, in this case, the dynamical Lamb effect is much more robust with respect to the dissipation in a qubit. The latter phenomenon can be treated as dissipation-assisted parametric amplification of vacuum, since a new channel of photon generation from vacuum opens due to the relaxation in a qubit.

We would like to stress that this striking increase of photon number is possible only when finite dissipation in a qubit is taken into account, since such a dissipation adds a new channel of photon generation from vacuum via qubit degrees of freedom. These results show that there are two competing processes in our system. The first one is due to counter-rotating processes, which
excite the qubit with simultaneous photon creation. The second one is a decay of the qubit excited state accompanied by oscillations due to excitation-number conserving processes. Which one prevails, depends on the character of a modulation. We also demonstrate that this competition can be described by the balance of two parameters which are nothing but two first Fourier components of a vacuum Rabi frequency as a function of time. The second regime is possible only for a strong driving, such that the coupling constant changes its sign during the modulation. A modulation of this sort seems to be possible for present or near-future technologies. Thus, we hope that the change of the behavior we predict here can be observed in experiments.

We also analyzed in a more detail the effect of cavity relaxation. We find that the difference between the two regimes is smeared out, since in both cases the stabilization is finally achieved, but at long enough times. Moreover, the nonzero cavity relaxation always leads to the enhancement of qubit excited state population at long times. Hence, by increasing this quantity, one can also increase the dynamical Lamb effect. This increase is stronger for those types of modulation which lead to the decay of this probability in the dissipation-free case. Thus, an increase of cavity relaxation provides an alternative method to enhance an effect. This method is of importance because it does not require the usage of sign-alternating modulations which can be technically difficult to implement.

The investigation of responses of quantum systems on nonadiabatic modulation of their parameters is of interest not only from the viewpoint of realization of various fundamental QED effects, but also for purposes of quantum computation. Indeed, high-speed gates can induce various nonstationary QED effects related to vacuum amplification and parametric generation of excitations from vacuum. Therefore, both the understanding and the control of such effects is of great importance.

\begin{acknowledgments}
The authors acknowledge useful comments by A. V. Ustinov, O. V. Astafiev, K. V. Shulga, E. S. Andrianov, S. V. Remizov, and L. V. Bork. This work was supported by RFBR (project no. 15-02-02128). Yu. E. L. was supported by the Program of Basic Research of HSE.
\end{acknowledgments}

\end{document}